\documentclass[journal=
jcisd8,
manuscript=article]{achemso}

\usepackage{chemformula} % Formula subscripts using \ch{}
\usepackage[T1]{fontenc} % Use modern font encodings
\usepackage[version=3]{mhchem}
\usepackage{array}
\usepackage{booktabs}
\usepackage{mathpazo}
\usepackage{listings}
\usepackage{lmodern}
\usepackage{microtype}
\usepackage{adjustbox}
\usepackage{amsmath}
\usepackage[labelformat=parens,labelsep=quad,skip=3pt]{caption}
\usepackage{graphicx}
\usepackage{float}
\usepackage{multirow}

 \SectionNumbersOn
\usepackage{hyperref}
\hypersetup{
    colorlinks=true,
    urlcolor=blue
}

\author{Artem Glova}
\affiliation{Department of Physics and Astronomy, The University of Western Ontario, 1151 Richmond Street, London, Ontario N6A\,3K7, Canada}
\email{adglova@proton.me}

\author{Mikko Karttunen}
\affiliation{Department of Physics and Astronomy, The University of Western Ontario, 1151 Richmond Street, London, Ontario N6A\,3K7, Canada}
\alsoaffiliation{Department of Chemistry,  The University of Western Ontario, 1151 Richmond Street, London, Ontario, N6A\,5B7, Canada}
\email{mkarttu@uwo.ca}
\title%[An \textsf{achemso} demo]
  {Learning glass transition temperatures via dimensionality reduction with data from computer simulations: Polymers as the pilot case
  \footnote{ Electronic supplementary information (ESI) available. See DOI: }}

\begin{document}


%%%%%%%%%%%%%%%%%% REMEMBER THIS
\begin{tocentry}
{\centering
    \includegraphics{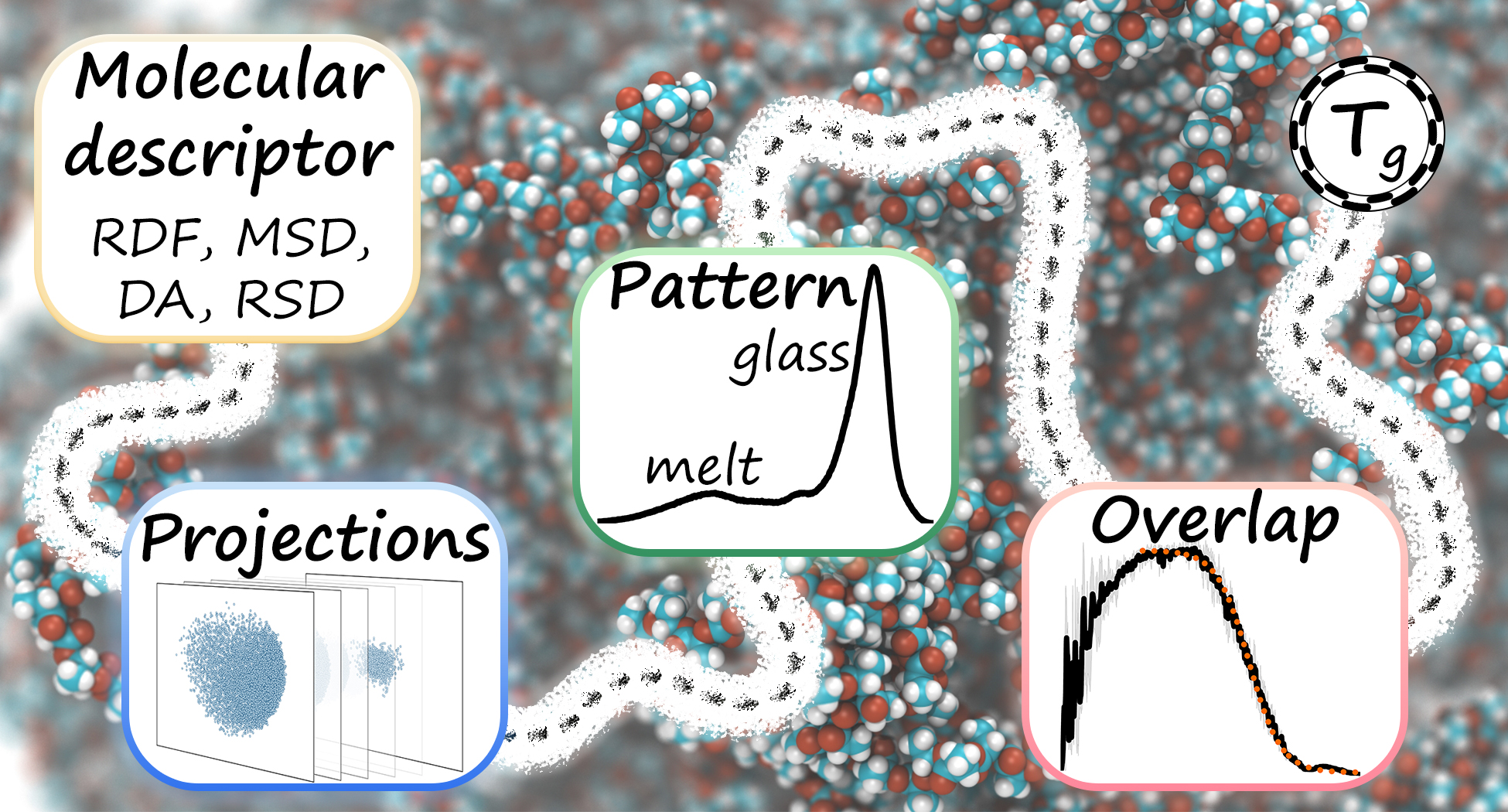}
\par
}
%%%%%%%%%%%%%%%%%%%%%%%%%%%%%
\end{tocentry}

%The surrounding frame is 9\,cm by 3.5\,cm, which is the maximum permitted for  \emph{Journal of the American Chemical Society} graphical table of content entries. The box will not resize if the content is too big: instead it will overflow the edge of the box.

%This box and the associated title will always be printed on a separate page at the end of the document.

%%%%%%%%%%%%%%%%%%%%%%%%%%%%%%%%%%%%%%%%%%%%%%%%%%%%%%%%%%%%%%%%%%%%%
\newpage 

\singlespacing

\begin{abstract}

Machine learning (ML) methods provide  advanced means for understanding inherent patterns within large and complex datasets. Here, we employ the principal component analysis (PCA) and the diffusion map (DM) techniques to evaluate the glass transition temperature ($T_\mathrm{g}$) from low-dimensional representations of all-atom molecular dynamic (MD) simulations of polylactide (PLA) and poly(3-hydroxybutyrate) (PHB). Four molecular descriptors were considered: radial distribution functions (RDFs), mean square displacements (MSDs), relative square displacements (RSDs), and dihedral angles (DAs). By applying a Gaussian Mixture Model (GMM) to analyze the PCA and DM projections, and by quantifying their log-likelihoods as a density-based metric, a distinct separation
%of log-likelihoods 
into two populations corresponding to melt and glass states was revealed. 
This separation enabled the $T_\mathrm{g}$ evaluation from a cooling-induced sharp increase in the overlap between log-likelihood distributions at different temperatures. $T_\mathrm{g}$ values derived from the RDF and MSD descriptors using DM closely matched the standard computer simulation-based dilatometric and dynamic $T_\mathrm{g}$ values for both PLA and PHB models. This was not the case for PCA. 
The DM-transformed DA and RSD data resulted in 
$T_\mathrm{g}$ values in agreement with experimental ones. Overall, the fusion of atomistic simulations and diffusion maps complemented with the Gaussian Mixture Models presents a promising framework for computing $T_\mathrm{g}$ and studying the glass transition in a unified way across various molecular descriptors for glass-forming materials.

\end{abstract}

\newpage


\section{Introduction}

Glass formers represent an abundant family of natural and manufactured materials composed of, e.g., polymeric, molecular, network, ionic, and metallic substances\cite{Lunkenheimer2023-eh}.
When cooled from a melt, they can fall out of equilibrium and develop heterogeneous, non-Arrhenius, and dramatically reduced dynamics\cite{Debenedetti2001-eq}. This makes them resemble solids but lacking long-range structural order. At the microscopic scale, approaching $T_\mathrm{g}$ can involve continuous kinetical constraints on cooperative particle rearrangements, which nonetheless remain facilitated due to localized motions confined within the region of their surrounding neighbours\cite{Stillinger1995-lz,Ngai2023-kd,Ghanekarade2023-av}. Gaining a rigorous understanding of the molecular mechanisms responsible for the glass transition
%phenomenon 
remains a grand challenge in materials science\cite{Biroli2013-qg,Ediger2012-tn}. From a practical perspective, $T_\mathrm{g}$ is a key characteristic in selecting a glass former for an application. Therefore, detection of the glass transition, and accurate measurement of $T_\mathrm{g}$ constitute fundamental steps in understanding this phenomenon, and in making application-oriented choices and advancements of glass-forming materials.

Experimentally, glass transition can be studied using various dilatometric, calorimetric, spectroscopic, microscopic and scattering methods, as well as dynamical mechanical and thermomechanical analyses\cite{Angell1995-me,Zheng2019-dg,Ruiz-Ruiz2023-rz,Robeson2007-fa,Roth2016-ni}. The canonical signature of $T_\mathrm{g}$ 
%employed for its evaluation 
in volume dilatometry measurements\cite{Roth2016-ni} is the change in slope of the temperature dependence of the sample volume. Similarly, $T_\mathrm{g}$ can be observed from neutron scattering experiments by identifying a kink in the mean square particle displacements versus temperature\cite{Angell1995-me}. Differential scanning calorimetry (DSC), 
enables $T_\mathrm{g}$ determination based on the step-like variation of heat capacity during a temperature scan\cite{Zheng2019-dg}. Experimental analyses are complicated by the purely microscopic nature of the glass transition governed by nontrivial many-body interatomic interactions and molecular relaxations that remain underexplored\cite{Ngai2023-kd}. In this regard, computer simulation methods allow an alternative approach\cite{Berthier2023-bd,Riniker2018-jx}.

Pioneering computational studies of glass formation date back about 60 years\cite{Alder_1960}. For the common and well-established method of $T_\mathrm{g}$ evaluation based on the density-temperature dependence $\rho(T)$, Carmona Esteva \textit{et al.}\cite{Carmona_Esteva2023-fc} have recently put forward an automatic protocol of kink identification, which solves an optimization problem assuming the existence of two linear domains related to the glass and melt states in $\rho(T)$ of ionic liquids. Patrone \textit{et al.}\cite{Patrone2016-hd} tested the applicability of a hyperbolic fit of $\rho(T)$. In simulations, however, 
%this dependence can experience 
two kinks have been reported previously for polyethylene\cite{Godey2018-ap}. Lin \textit{et al.}\cite{Lin2021-rm}
found that the $\rho(T)$ curves for organic polycyclic molecular glass formers can have multiple regions with different slopes, and as a result they suggested a novel procedure to determine the target temperature range for a bilinear fit. A common alternative to dilatometric $T_\mathrm{g}$ analysis is to calculate the atomic mean square displacements $\langle r^2 \rangle$ in analogy to experiments\cite{Angell1995-me}.
When it comes to polymers, for example, Morita \textit{et al.}\cite{Morita2006-ka}
predicted dynamic $T_\mathrm{g}$ of films from a kink of segment $\langle r^2(\Delta t^*) \rangle$
%<r2(Δt*)> 
measured at short time intervals $\Delta t^*$
%Δt* 
as a function of temperature. In later studies, Arrhenius-like plots have been used, such as 
$\langle r^2(\Delta t^*) \rangle (1/T)$
%<r2(Δt*)>(1/T), 
as well as 
$\ln (1/r^2(\Delta t^*))$
%ln(1/r2(Δt*)) 
or 
$(d(r^2(\Delta t))/d \Delta t)_{\Delta t = \Delta t^*} $
%(d(r2(Δt))/dΔt)Δt=Δt* 
versus $1000/T$\cite{Glova2018-ur,Sun2022-kb,Tanis2023-ox}.
Zhou and Milner\cite{Zhou2017-hb}
suggested a sixth-order polynomial fit of
$\langle r^2(\Delta t^*) \rangle (T)$
%<r2(Δt*)>(T) 
to extract $T_\mathrm{g}$ shifts within polymer films. These studies exemplify the fact that several approaches to dynamic $T_\mathrm{g}$ alone can be found in the literature on polymeric glass formers. More generally, different other metrics for the $T_\mathrm{g}$ estimation from computer simulations have been employed, including but not limited to radial distribution functions and dihedral angles\cite{Kolotova2015-en,Godey2019-fb,Hung2019-ry,Ojovan2020-mb,Alzate-Vargas2020-es,Jin2022-dc,Tang2022-wi}.

Importantly, most methods to determine the $T_\mathrm{g}$ values from computer simulations operate with averaged microscopic and macroscopic quantities. The output of a simulation represents a high-dimensional dataset, with complexity being governed by the number of interacting atoms and the length of their trajectories over time. 
The typical time step is of the order of a few femtoseconds
meaning, on the one hand, that simulations can easily produce terabytes of trajectory data,
and on the other hand, that the $T_\mathrm{g}$ evaluation methods do not fully utilize microscopic details available in pristine simulation outcomes.

Dimensionality reduction techniques represent a powerful machine learning technology to solve the above issue\cite{Glielmo2021-vg}.
Given the microscopic conundrum of the glass transition and the corresponding lack of its general definition, the prospect of starting with the time evolution of atomic ensembles and establishing molecular patterns of the glass transition from its low-dimensional representation to predict “machine learned” $T_\mathrm{g}$ values make the application of these techniques intriguing. Empirically, simulations do capture the glass transition process, and therefore may contain those patterns. Although the number of papers per month on artificial intelligence and machine learning has grown exponentially over the last several years\cite{Krenn2023-ds}, only a few studies have focused on dimensionality reductions for $T_\mathrm{g}$ estimation\cite{Iwaoka2017-jo,Banerjee2023-xx,Banerjee2023-wh}. Iwaoka and Takano\cite{Iwaoka2017-jo} simulated coarse-grained oligomeric chains and employed principal component analysis (PCA) to time-evolving coordinates for each bead upon cooling. Assuming Gaussian statistics for oligomers, they found it possible to obtain a $T_\mathrm{g}$ value using eigenvalues of covariance matrices for all beads, i.e., their variance contributions, at various temperatures. In a similar fashion, based on PCA of intramolecular distances over time for a bead-spring model Banerjee \textit{et al.}\cite{Banerjee2023-xx} suggested that a $T_\mathrm{g}$ value can be predicted from a non-monotonic change in the variance explained by the dominant principal component (PC) or in the participation ratio for the first 25 dominant PCs accounting for at least 80\% of total variance in data. They also proposed another method, where the DBSCAN (Density-Based Spatial Clustering of Applications with Noise) algorithm\cite{Schubert2017-wt} was applied to automatically cluster PCA projections combined from all temperatures for individual chains and determine the $T_\mathrm{g}$; note that DBSCAN hyperparameters were tuned using the V-measure score and explicitly defined reference temperature ranges of the glass and melt states. In their recent study, Banerjee \textit{et al.} extended their methods to realistic acrylic oligomers and reported that $T_\mathrm{g}$ values from the clustering of PCA data can be markedly (up to about 90 degrees) greater than dilatometric $T_\mathrm{g}$ values with the discrepancy depending on the oligomer types. They also tested the \texttt{cc\_analysis} method\cite{Diederichs2017}, a nonlinear dimensionality reduction technique, and found that this method consistently results in $T_\mathrm{g}$ values slightly greater than those from PCA\cite{Banerjee2023-wh}.

Overall, dimensionality reduction techniques can enable the formulation of unified frameworks for learning the glass transition from computer simulations. In the case of realistic models, 
%ML 
$T_\mathrm{g}$ values may mismatch with the dilatometric ones, possibly due to the choice of molecular descriptor(s). At an equal extent, the choice of dimensionality reduction method can also impact resulting $T_\mathrm{g}$ values. 
%
%Previously used\cite{Iwaoka2017-jo,Banerjee2023-wh,Banerjee2023-xx}
PCA can be performed via spectral decomposition of the covariance matrix (or correlation matrix), where PCs represent linear projections of data onto a new set of orthogonal dominant directions with maximized variance\cite{Jolliffe2016-qi}. The performance of PCA may degrade when applied to non-Gaussian distributed data, as the covariance matrix may end up describing the data structure inaccurately and missing possible nonlinear relationships between the variables\cite{Lim2009PRINCIPALCC,Stamati2010-wn,Shlens2014-vw,David2014-fz}. The coordinates of many molecular processes involve large-scale atomic motions and, for enzymatic reactions and protein folding, are expected to reside on low-dimensional nonlinear hypersurfaces (manifolds)\cite{Stamati2010-wn,Das2006-vp,Antoniou2011-lq},
where PCA will yield only their linear approximations in terms of variance distribution. 

Ceriotti\cite{Ceriotti2019-sx}
highlighted that “when working with actual data rather than synthetic data, there is no ‘ground truth’ in terms of what partitioning of the data, or low-dimensional representation, is performing best”. For the glass transition phenomenon, exploring how dimensionality reduction techniques perform for different molecular descriptors to accurately measure its fundamental metric - $T_\mathrm{g}$ - remains largely an open question. We focus on this question in the present work.

To this end, we carry out a comparative analysis of PCA and the diffusion map methods for $T_\mathrm{g}$ measurements based on atomistic MD simulations of polymeric glass formers polylactide and poly(3-hydroxybutyrate), which are well-studied in experiments and simulations.\cite{Glova2018-ur,Farah2016-so,Sudesh2000-ya}
DM is a general-purpose nonlinear dimensionality reduction technique with robust mathematical basis for running various diffusion processes on data to reveal the geometry of its low-dimensional manifold\cite{Coifman2005-lq,Coifman2006-sm}. Compared to the notion of variance in PCA, DM allows one to find the dominant time scales of the stochastic paths within the data. DM has been successfully used to study the conformational properties of alkane chains and peptides\cite{Ferguson2010-fu,Ferguson2010-ls,Mansbach2015,Mansbach2023,Trstanova2020-fn}, the self-assembly of polycyclic aromatic compounds\cite{Wang2018-th}, the dynamics of Lennard-Jones systems\cite{Coifman2008-vi}, single-cell differentiation trajectories\cite{Haghverdi2015-un}, and the reaction dynamics in hydrogen combustion systems.\cite{Ko2023-ml} In this work, we explore DM and PCA projections of various molecular descriptors, with the primary emphasis on unraveling the detailed changes in their structures as a function of temperature. This involves the need to determine an approach for selecting the number of components for these two different dimensionality reduction methods, choosing a metric to describe the structure of representations, analyzing this metric to pinpoint the transition temperature, and validating our 
%ML 
$T_\mathrm{g}$ estimation framework.

%%%%%%%%%%%%%%%%%%%%%%%%%%%%%%%%%%%%%%%%%%

\section{Methods}
\label{sec:methods}

%% vvvvvvvvvvv  THIS SECTION IS OK

\subsection{Simulation details}
\label{sec:simu_details}

\begin{figure}[tbhp]%[H]%[tbhp]
\centering
\includegraphics[width=0.6\linewidth]{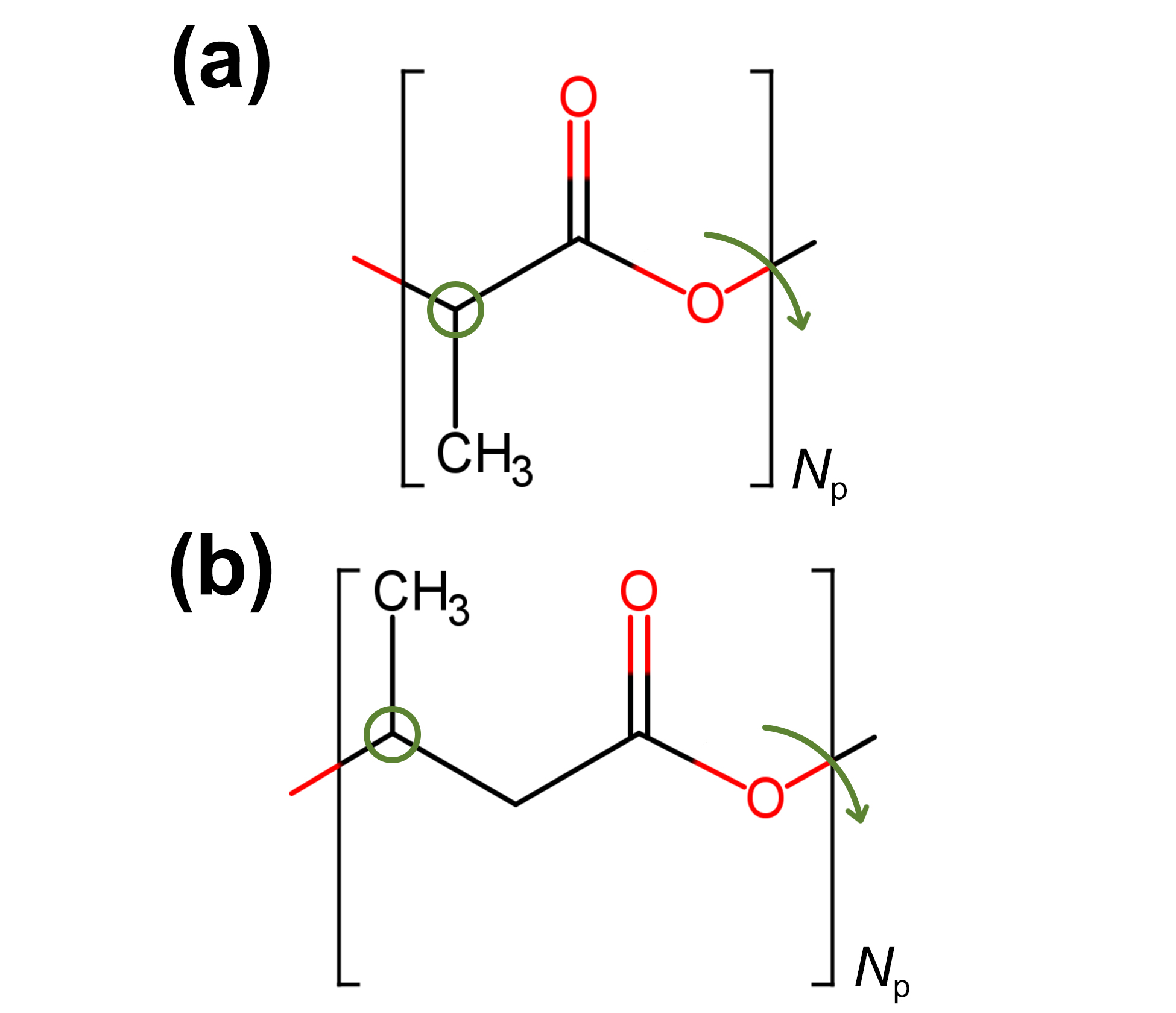}
\caption{Repeating units of (a) PLA and (b) PHB. Circles and arrows indicate the chiral carbon atom and the dihedral angle within the chain backbone used to construct molecular descriptors. The degree of polymerization, $N_\mathrm{p}$, was 150 for both PLA and PHB. The green arrows also show the dihedral angles used for computing autocorrelation functions.}
\label{fig:pla_phb_structure}
\end{figure}

The Gromacs 2022.3 package\cite{Abraham2015-zu} was used for all MD simulations of PLA and PHB (Figure~\ref{fig:pla_phb_structure}). 
We adopted the simulation methodology previously validated for the description of the physical properties of these polymers, such as the persistence length and glass transition temperature.\cite{Glova2018-ur,Glova2020-kg}.
The general Amber force field (GAFF)\cite{Wang2004-ho}
was employed to build atomistic models of PLA and PHB. The partial charges were derived using the standard HF/6-31G$^*$(RESP) method based on tetramers, facilitated by the pyRED program version SEP-2022 via the R.E.D. Server Development 2\cite{Vanquelef2011-kc,Dupradeau2010-bt}.
Bonded and van der Waals interactions were parameterized through the ACPYPE program version 2022.7.21\cite{Sousa_da_Silva2012-kd}.

The degree of polymerization was set to 150, resulting in molecular weights of about 11\,kg/mol for PLA and 13\,kg/mol for PHB. Note that molecular weights exceeding 10\,kg/mol have been found to be sufficient for reaching the plateau values in the Fox–Flory plot for the glass transition temperatures of PLA\cite{Zhang2022-qg}, i.e., this selection of chain lengths corresponds to the polymer regime. The isomeric content was typical for the two polymers: PLA and PHB chains consisted of S and R isomers, respectively\cite{Garlotta2001-ze,Sudesh2000-ya}. The initial coordinate files of the chains in a stretched conformation were generated using the CHARMM-GUI Polymer Builder.\cite{Choi2021-zp}

To construct model systems, each polymer chain was placed inside a cubic box with a minimal distance of 1\,nm from its edges. The systems were coupled with the Berendsen thermostat\cite{Berendsen1984-db} with a time constant of 0.1\,ps to maintain the temperature $T$ at 550\,K. This temperature exceeds the melting points of PLA and PHB by about 100 degrees\cite{Van_de_Velde2002-vp}, allowing the systems to be in the melt state. Initially, a short simulation for 25\,ps was carried out to slightly decrease chain dimensions. Then, 50 chains of PLA or PHB were randomly positioned within a cubic box with edge lengths of about 25\,nm. This arrangement yielded a total of 67\,650 atoms for PLA and 90\,150 atoms for PHB. After that, the Berendsen barostat\cite{Berendsen1984-db} with a time constant of 1\,ps and a pressure ($P$) of 50\,bar was applied to conduct isotropic compression for 5\,ns; the Berendsen thermostat and barostat should not be used for production simulations, but due to their stability, they are very helpful when setting up complex systems. 

Next, we switched the systems to the Nos\'e–Hoover thermostat\cite{Nose1984-cm,Nose1984-so,Hoover1985-kn} with a time constant of 1\,ps and the Parrinello–Rahman barostat\cite{Parrinello1981-zl} with a time constant of 5\,ps and $P$\,=\,1\,bar. PLA was simulated for 3.5\,$\mu$s and PHB for 1.5\,$\mu$s. Analysis of the time dependence of the radius of gyration $R_\mathrm{g}$ showed that the chains reached sizes of $3.0 \pm 0.1$\,nm for PLA and $3.4 \pm 0.2$\,nm for PHB during the initial 300\,ns of simulations, see 
Figure~S1. 
These equilibrium $R_\mathrm{g}$ values are in agreement with previous results\cite{Glova2018-ur}, justifying the models and protocol. The last 1\,$\mu$s of simulations were utilized to collect three independent configurations spaced apart by intervals of 500\,ns, exceeding the time of about 300–400\,ns required for monomers to displace by a distance comparable to the chain $R_\mathrm{g}$\cite{Glova2018-ur}.

The above configurations were used as inputs for the simulations of cooling, where the temperature was varied from 550\,K to 1\,K with a step of 1\,K every 0.1\,ns (resulting in 55\,ns per scan). Such a cooling rate is typical for atomistic MD simulations of glass formers\cite{Lin2021-rm,Lyulin2014-oy,Li2017-hq}. These cooling rate simulations were utilized to estimate dilatometric and dynamic $T_\mathrm{g}$ values. To this end, we employed the standard approaches based on the bilinear fit of temperature dependence for the volume and mean square displacements\cite{Morita2006-ka,Baljon2010-kq}
of chiral carbon atoms over 20\,ps, as illustrated in 
Figure~S2. 
Moreover, the cooling rate simulations provided inputs for the ML analysis, as will be described in Section~\ref{sec:mol_descriptors}. 

Periodic boundary conditions were applied in all three dimensions, and all bonds were constrained using the P-LINCS algorithm\cite{Hess2008-oc}. The electrostatic interactions were computed with the particle-mesh Ewald method~\cite{Darden1993-qr,Essmann1995-sg}. Non-bonded interactions were cut off at 0.9\,nm, beyond which a correction to the van der Waals energies were added\cite{Wang2011-cj}.
The integration time step was set to 1\,fs. 

The trajectory writing step (WS; the frequency of writing output data) may matter for the $T_\mathrm{g}$ analysis as it determines the quality of sampling and, consequently, the resolution for monitoring relaxation. Following 
Takeuchi and Okazaki\cite{Takeuchi1990-yh}, we analyzed the autocorrelation functions of the dihedral angles (see Figure~\ref{fig:pla_phb_structure}) based on additional MD simulations of PLA and PHB at $T$ = 550\,K for 0.1\,ns, with data written every 1\,fs. 
Figure~S3 
shows that the spectrum of dihedral angle relaxation is complex and involves several processes. The fastest process completes in about 30\,fs. At the same time, the correlation time for the dihedral angles decreases by a factor of $e$ in about 2\,ps, being a possible upper limit for the WS enabling reliable sampling. To verify these assumptions and to motivate the choice for WS, we wrote out data every 5\,fs and considered four writing steps for PLA: 10\,fs, 25\,fs, 100\,fs, and 1000\,fs, see Section~\ref{sec:writing_step}. In other words, we analysed data within the time scales of the fastest relaxation process (WS = 10\,fs and 25\,fs), as well as the longer time scales below the upper limit (WS = 100\,fs and 1000\,fs). In the case of PHB, WS was set to 25\,fs. Note that, considering the chosen WS, each cooling scan required about 1.7\,TB and 700\,GB of disc space to store PLA and PHB trajectories, respectively.

%% THIS SUBSECTION IS OK vvvvvvvvvvvvvvvvvvvvvvvvvvvvvvvvvvvvv

\subsection{Molecular descriptors}
\label{sec:mol_descriptors}

Having performed the cooling rate simulations, we measured radial distribution functions, mean square displacements, relative square displacements, and dihedral angles. To generate the descriptor datasets, we utilized the MDAnalysis\cite{Michaud-Agrawal2011-td,Gowers2016-za} and NumPy\cite{Harris2020-gz} packages.

In order to reduce the computational burden, we focused on collecting descriptor datasets based on the MD trajectories for the chiral carbon atoms and dihedral angles within the chain backbone as depicted in Figure~\ref{fig:pla_phb_structure}. However, we further expanded our study to include other possible types of atoms and dihedral angles, see
Figure~S4 
for their definitions, and Section~\ref{sec:no_observables} for the corresponding discussions.

First, we computed RDFs ($g_i(r)$) for the chiral atoms. The dataset comprised $N = 7\,400$ chiral atoms (samples) for both PLA and PHB, and did not include the chain ends. To avoid probing possible local effects on the resulting $T_\mathrm{g}$ stimate\cite{Glova2018-ur}, the $g_i(r)$ were cut off at 2\,nm when they approached unity, see
Figure~S5a. 
Consequently, each atom was represented by a feature vector of 200 RDF values, resulting in an input array of shape (7\,400, 200) for a given $T$. In cases involving several atom types, 
(Figure~S4), 
calculations followed a similar approach and considered the distribution of specific atoms near chiral carbon atoms.

In addition to their previous application for estimating $T_\mathrm{g}$,\cite{Kolotova2015-en,Ojovan2020-mb} local structures expressed by RDFs have been  utilized for predicting structural rearrangements in glass formers\cite{Schoenholz2016-fv,Ma2019-te}. It is worth noting that the RDF serves as a truly structural descriptor and is directly proportional to the system volume. Hence, its utilization and comparison with the dilatometric $T_\mathrm{g}$ analysis 
(Figure~S2a) 
offer a viable means to judge the performance of dimensionality reduction techniques. 

MSDs present another possible way for assessment, as they can be analyzed using both the standard approach 
(Figure~S2b) 
and dimensionality reduction techniques. For a chiral atom $i$, MSDs 
were calculated as $\langle r_i^2(\Delta t) \rangle = \langle | \vec{r}_i(t + \Delta t) - \vec{r}_i(t) |^2 \rangle$, where $\vec{r}_i (t)$ and $\vec{r}_i (t + \Delta t)$ are the positions of the $i^\mathrm{th}$ atom at times $t$ and $t + \Delta t$, respectively, and $\langle \cdots \rangle$ denotes an average over the simulation time. We set the maximal time interval to $\Delta t_\mathrm{max} = 20$\,ps and used 20\% of available frames for the MSDs, as the quality of time averages diminishes with increasing $\Delta t$, see 
Figure~S5b. 
This choice also explains the selection of the time interval for the standard analyses 
(Figure~S2b). 
All chiral atoms were analyzed. Given the chosen WS of 25\,fs, each MSD feature vector contained 800 values. Thus, the input MSD descriptor represented an array with a shape of (7\,500, 800). For the extended dataset with additional atoms
(Figure~S4), 
the array shape was (30\,000, 800) for PLA and (37\,500, 800) for PHB.

In turn, dihedral angles and relative square displacements can provide valuable insights into the systems, offering pristine MD data for analysis. To account for the circular nature of the dihedral angles and to reduce possible projection errors in DM and PCA\cite{Sittel2017-ur}, DA values calculated from these trajectories were shifted by $\pi$ for both PLA and PHB, as shown in
Figure~S6. 
The RSDs were evaluated based on the positions of the chiral atom $i$ at times $t$ relative to its position at the initial time $t = 0$\,ns using $r_i^2(t) = | \vec{r}_i(t) - \vec{r}_i(0) |^2$, where $i$ represents all 7\,500 chiral atoms in the PLA and PHB systems. While MSDs describe averaged dynamics, RSDs track the evolution of atomic positions with respect to their starting coordinates. The trajectories for both DAs and RSDs generated at intervals of 25\,fs consisted of 4\,000 frames, which defined the length of the corresponding feature vectors. The input arrays for the DA and RSD descriptors had shapes of (7\,450, 4\,000) and (7\,500, 4\,000), respectively. Extended datasets of DAs were stored in arrays of shape (22\,350, 4\,000) for PLA and (29\,850, 4\,000) for PHB, while RSDs were represented by arrays with shapes (30\,000, 4\,000) for PLA and (37\,500, 4\,000) for PHB.
 
\subsection{ML methods}
\label{sec:ml_methods}

%%% This subsection is OK vvvvvvvvvvvvvvvvvvvvvvvvv

In this work, we transformed the descriptor data using DM and PCA. The latter technique was selected as a reference based on previous studies\cite{Iwaoka2017-jo,Banerjee2023-wh,Banerjee2023-xx}.
Following Banerjee \textit{et al.}~\cite{Banerjee2023-wh,Banerjee2023-xx}
the descriptor data was standardized by centering each feature vector of a sample, and dividing it by its standard deviation prior to employing DM and PCA. PCA was carried out by means of the scikit-learn package~\cite{scikit-learn}.

DM was performed using the pyDiffMap package\cite{Banisch2017-th,Banisch2020-gc}. 
It involves the construction of a neighborhood graph on the input data based on a Gaussian kernel for pairwise distances between samples normalized several times to yield a generator of a Markov chain\cite{Coifman2005-lq,Coifman2006-sm}.
This implies setting hyperparameters such as the kernel bandwidth $\varepsilon$, the number of nearest neighbor points $k$, and a normalization constant $\alpha \in [0, 1]$. $\varepsilon$ is a length scale hyperparameter defining the dynamical proximity between the samples, $k$ determines the number of samples considered in graph construction, and $\alpha$ controls the influence of sampling density of data on diffusion. Given our simulation setup, where a small temperature step of 1\,K was used upon cooling (Section~\ref{sec:mol_descriptors}), we can assume that the systems do not transition from a melt to a glass state within a single cooling step. To incorporate this assumption in DM and to get a global representation of a given state, we treated most samples as dynamically similar, and set $\varepsilon$ and $k$ equal to the variances of the pairwise distances between the samples and the number of input samples, respectively. The pairwise distances were computed using the SciPy package\cite{scipy}. For the thus constructed kernel, no dependency of $T_\mathrm{g}$ analysis on $\alpha$ was found (data not shown), and we present the outcomes obtained at $\alpha =0$. Note that both DM and PCA require the selection of the number of dominant directions in the data. This will be discussed in Sections~\ref{sec:dimensionality_projections} and \ref{sec:proj_dimensionality}.

DM and PCA projections served as inputs for other ML algorithms implemented in the scikit-learn package. Specifically, Gaussian Mixture Models were employed to derive log-likelihoods $L_\mathrm{GMM}$, which describe the logarithm of the likelihood of observing a projected data point (i.e., sample) within a multivariate Gaussian mixture\cite{McLachlan2019, Bishop_undated-lj}.
The number of mixture components approximating the projections was set to 1 because a single cluster was observed for all the descriptors transformed by both DM and PCA, see Section~\ref{sec:visual}. Moreover, the OPTICS algorithm was employed to compute the so-called core distances of points\cite{Ankerst1999-ro}. Here, the core distance for a point describes the minimum distance required to encapsulate all other points of a projection. Finally, we computed the mean square distances for each point to all its neighbors using the nearest neighbor search. These three density-based metrics were utilized to evaluate the structure of DM and PCA projections. The main discussion is based on the log-likelihoods, while the impact of these metrics on the $T_\mathrm{g}$ estimation will be analyzed in Section~\ref{sec:density_based_metrics}.

%%%% ^^^^^^^^^^^^^^^^^^^^^^^

\section{Results}

%% THIS SUBSECTION IS OK AFTER THE COMMENTS HAVE BEEN ADDRESSED vvvvv

\subsection{Dimensionality of projections}
\label{sec:dimensionality_projections}

One of the key steps in dimensionality reduction is determining the number of dominant eigenvectors and, correspondingly, the dimensionality of the projection. In PCA, this can be achieved by considering the fraction of the total variance explained by a certain number of eigenvectors. However, the decision regarding how many eigenvectors to keep is often arbitrary, typically aiming to capture around 95\%-98\% of the dataset’s variance\cite{Glielmo2021-vg}.

This notion of variance is not applicable to DMs, where eigenvalues describe the time scales of diffusion in data. In this regard, to compare the DM and PCA outcomes, one needs a common approach for determining the projection’s dimensionality. To this end, we define spectral gaps as 
\begin{equation}
\sigma_k = \lambda_{k-1} - \lambda_k ,
\label{eq:sigma_k}   
\end{equation}
 where $\lambda_i$ denotes the eigenvalue of the $i^\mathrm{th}$ eigenvector. For this analysis, we focus on the maximum temperature $T$ = 550\,K, which corresponds to a reference melt state where the system can be considered as equilibrated. We collected $\sigma_k$ from three independent simulations and then averaged them to enhance the quality of the analysis.

\begin{figure}[tbhp]%[H]%[tbhp]
\centering
\includegraphics[width=1.0\columnwidth]{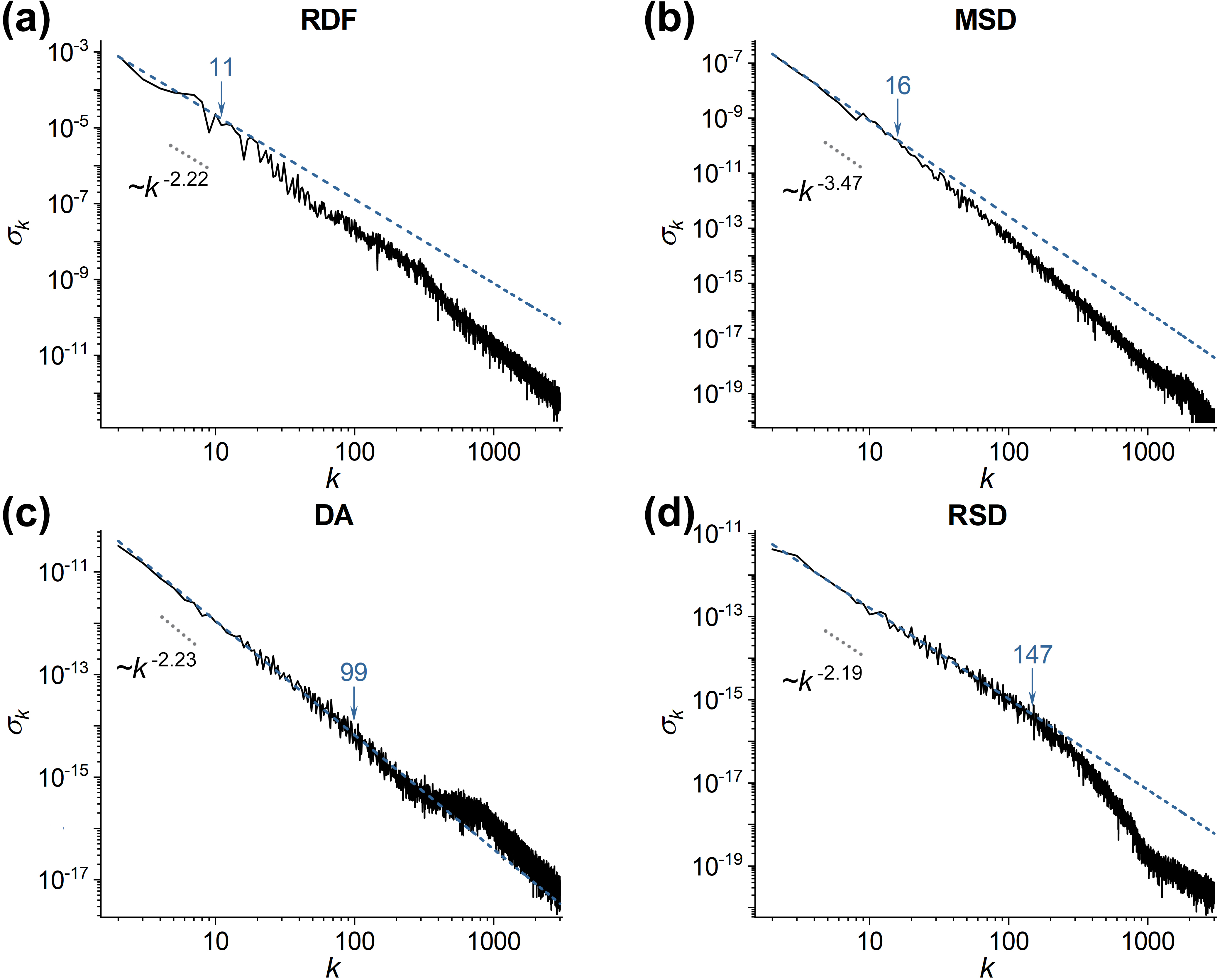}
\caption{Double logarithmic plots of spectral gaps $\sigma_k$ (Equation~\ref{eq:sigma_k}) obtained using DM for the (a) RDF, (b) MSD, (c) DA, and (d) RSD data of PLA at $T = 550$\,K. The dashed and dotted lines indicate fits of the initial decrease in $\sigma_k$ and their power laws, respectively. Note that the fitting range 
%$k\string0$ for these fits 
was determined based on the $R^2$-value of $\geq 0.95$ between the original $\sigma_k$-curve and its fit $\sigma_k^\mathrm{fit}$, while the number of dominant eigenvectors to consider was chosen using the criterion $\sigma_r < 0.2$, 
see Equation~\ref{eq:sigma_r}.
%where $\sigma_r = \sigma^\mathrm{fit}/\sigma_k - 1$. 
Arrows show the number of eigenvectors for the different descriptors.
}
\label{fig:spectral_gaps_DM}
\end{figure}

The $\sigma_k$ values obtained for different descriptors in the case of DM are shown using a log-log plot in Figure~\ref{fig:spectral_gaps_DM}. Although the shapes of the $\sigma_k$-curves in the log-log plot are complex, a power law decrease was observed for the first $\sigma_k$ values for all the descriptors, Figure~\ref{fig:spectral_gaps_DM}. Note that similar shapes of $\sigma_k$-curves were obtained for PCA, see Figure~S7. The MSD data, however, is an exception as the power law decrease is followed by a different non-linear trend in PCA rather than a steeper power law decline as observed in DM, see 
Figures~\ref{fig:spectral_gaps_DM}b~and~S7b. 

Based on the above analysis, we assume that the deviation from the initial power law for $\sigma_k$ may serve for determining the target number of eigenvectors to keep. To quantitatively characterize the deviation, the following procedure is developed: First, we consider $\sigma_k$ fragments of varying lengths starting from the origin and obtained their linear approximations $\sigma_k^\mathrm{fit}$ on the logarithmic scale using 
\begin{equation}
\log \sigma_k^\mathrm{fit} = A + B\log k ,
\label{eq:sigma_k_fit}   
\end{equation}
where $A$ and $B$ are fitting constants. In analogy to Ref.~\cite{Lin2021-rm}, the coefficient of determination ($R^2$) between $\sigma_k$ and $\sigma_k^\mathrm{fit}$ was computed. To this end, the \texttt{r2\_score} function of the scikit-learn package was employed. Then, we find the fitting range
%$k\string0$ 
as the number of $k$-values, where $R^2 \geq 0.95$. The resulting $\sigma^\mathrm{fit}$ (Figures~\ref{fig:spectral_gaps_DM}~and~S7)
based on this range was used to calculate the ratio 
\begin{equation}
\sigma_r = \sigma^\mathrm{fit}/\sigma_k - 1 .
\label{eq:sigma_r}    
\end{equation}
The target number of eigenvectors is defined using $\sigma_r < 0.2$ as a criterion, which was found reasonable based on a visual analysis of the outcomes (Figure~\ref{fig:spectral_gaps_DM}). Note that a moving average was applied to $\sigma_k$ and $\sigma_r$ during the calculations. 

The above procedure was used to establish the projection dimensionality for various descriptors of PLA and PHB, see 
Table~S1. 
The RDF and MSD descriptors, being averaged characteristics by definition, have much lower intrinsic dimensionality compared to more pristine data from the DA and RSD descriptors.
It is also noteworthy that both reduction types lead to an almost equal projection dimensionality for a given descriptor. Furthermore, the exponents of the power laws in the initial regime from DM and PCA are close to each other for the RDF and MSD data, and identical for the DA and RSD data, see 
Figures~\ref{fig:spectral_gaps_DM} and S7.
These observations indicate that DM and PCA identify a similar number of dominant directions with comparable relationships in terms of explained variances and diffusion time scales, respectively. Note that the choice of projection dimensionality could be crucial for obtaining a reliable estimate for $T_\mathrm{g}$. We elaborate this point in Section~\ref{sec:proj_dimensionality} below.

\subsection{Impact of temperature on DM and PCA projections: visual analysis}
\label{sec:visual}

\begin{figure}[tbhp]%[H]%[tbhp]
\centering
\includegraphics[width=1.0\columnwidth]{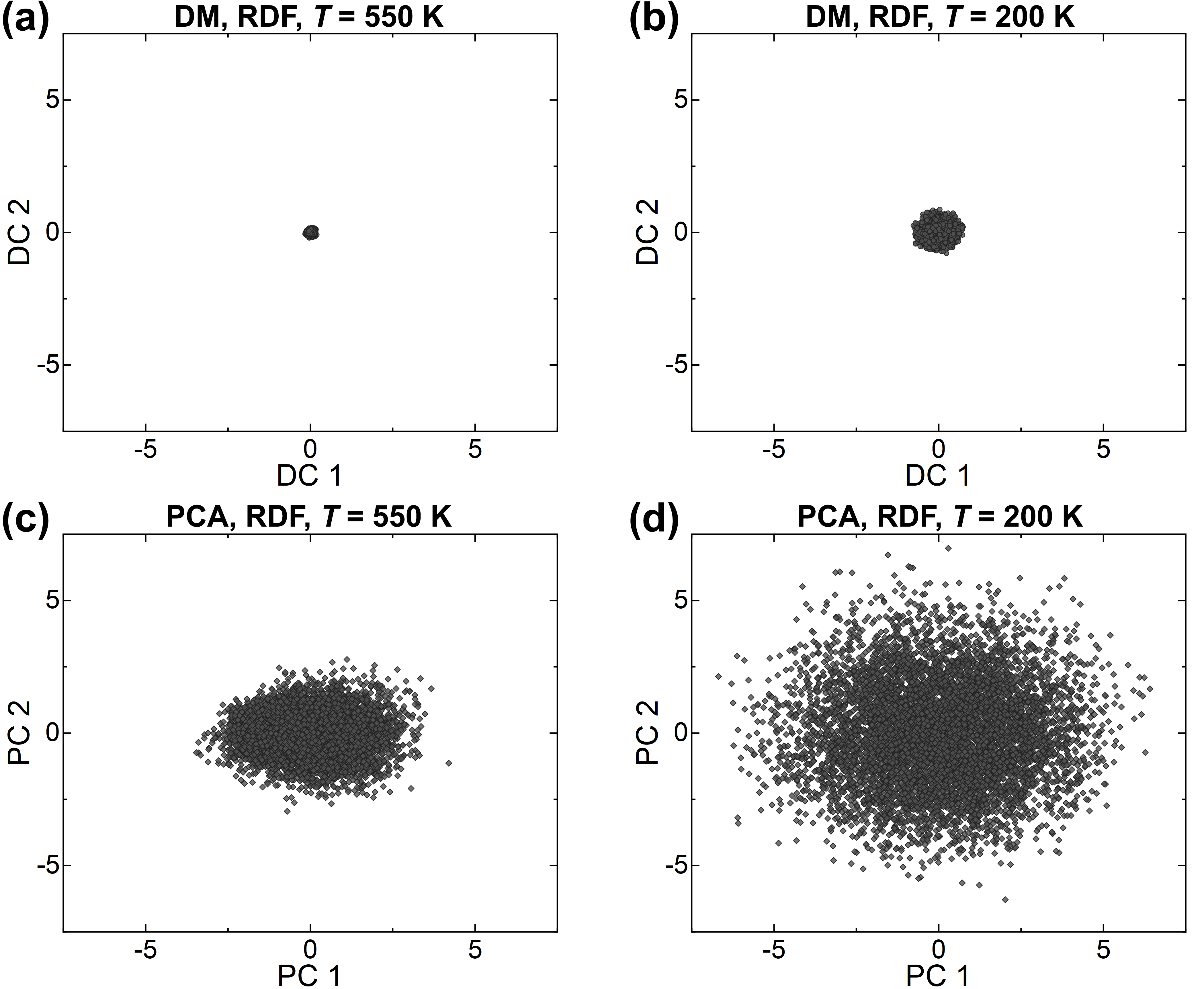}
\caption{(a, b) DM and (c, d) PCA projections of the RDF data onto the first two dominant eigenvectors for PLA at the temperature of (a, c) 550\,K and (b, d) 200\,K.
}
\label{fig:dm_and_pca_projections_of_rdf}
\end{figure}

As discussed in the Introduction and Section~\ref{sec:mol_descriptors},
simulations can successfully capture the glass transition. Hence, one can anticipate the presence of a hidden glass transition pattern within molecular trajectories. To unveil this pattern, we first focus on the basic results of dimensionality reduction.
%: DM and PCA projections. 
%%
%%^^^^^^^^^^^^^^^^^^^^^^^^^^^^^^^^^^^^^^^^^^^^^^^
Figure~\ref{fig:dm_and_pca_projections_of_rdf} shows the projections of the RDF data onto the first two dominant eigenvectors from DM and PCA at $T$ = 550\,K and 200\,K corresponding to the melt and glass states 
(see Figure~S2), 
respectively, for PLA cooled after 
3.5\,$\mu$s. Projections for the other descriptors, as well as for the second polymer, PHB, lead to similar observations, see 
Figures~S8-S14.

%^^^^^^^^^^^^^^^^^^^^^^^

%

As Figure~\ref{fig:dm_and_pca_projections_of_rdf} shows, the DM and PCA reductions of RDF indicate a single cluster. The size and shape depend on temperature and reduction type. Namely, for both DM and PCA the cluster appears as more compact at higher $T$. This may be attributed to the RDF curves having smoother and more similar shapes among different atoms in the melt as opposed to the glass state. Figure~\ref{fig:dm_and_pca_projections_of_rdf} also shows that at a given $T$, DM leads to a smaller cluster compared to PCA. Regarding the MSD data, the temperature and reduction type affect projection sizes in a similar manner, see 
Figure~S8. 

%% ^^^^^^^^^^^^^ ok

The observed temperature-dependent changes are in line with the recent results of Banerjee \textit{et al.}\cite{Banerjee2023-wh} as reported in Figure~S5 of their paper, where an increase in cluster sizes with decreasing temperature was also seen for PCA reductions of trajectories of intramolecular distances. Contrary to RDF and MSD, however, the effect of temperature on cluster sizes appears to be opposite for the DA and RSD descriptors, see 
Figures~S9~and~S10:
the clusters become less compact as temperature increases.
%are less compact at high $T$ than at low $T$. 
The larger clusters observed in the projected DA and RSD data for the melt state probably reflect increased mobility and greater variability among the trajectories of different atoms and dihedral angles.

%%%vvvvvvvvvvv CHECK 

Compared to PCA, DM of DA data yields a smaller cluster at a given $T$ (Figure~S9), similarly to RDF and MSD. For RSD, at high $T$ the cluster from DM appears larger than that from PCA, and vice versa at low $T$, so the role of reduction type can be a complex matter (Figure~S10).

As for the cluster shape, the RDF data transformed using DM exhibits nearly circular scattering of points, which remains practically independent of temperature, see Figures~\ref{fig:dm_and_pca_projections_of_rdf}a,b. A similar shape is observed for the MSD data at high $T$, while cooling results in the appearance of a sparse tail of points, see 
Figures~S8a,b. 

The DM transformation of DA at high $T$ leads to an almost round cluster with an inhomogeneous sparse periphery, which densifies upon cooling, see 
Figures~S9a,b. 
For the RSD data, DM projections resemble a round cluster, which has a dense domain on one side at high $T$ and becomes smaller and almost uniform at low $T$, see 
Figure~S10a,b. 
%The transition from DM to PCA leads to slightly different pictures: 
Compared to DM, PCA projections display a more oblate shape, see 
Figures~\ref{fig:dm_and_pca_projections_of_rdf} and S8-S10. Such a shape may reflect the presence of linear correlations between samples and suggest that PCA captures them.
Notably, overlaying the PCA projections on top of the DM projections accounting for different scales shows that the points scatter almost identically at both high and low $T$ with deviations depending on temperature and indexes of DC or PC, see 
Figures~S15-S18. 
This possibly means that principal components and diffusion coordinates (DCs) not only possess similar relative contributions to variance and diffusion time scales (Section~\ref{sec:dimensionality_projections}), respectively, but can also point to similar dominant directions in the data. Although the construction of DM involves the application of the diffusion kernel and normalization steps, and thereby forbids straightforward comparisons of DC and PC scales, the observed difference in scales may be partly attributed to various meanings of distances in PCA and DM. While it expresses the geometric proximity in PCA, the distance for DM projections describes the sum of stochastic paths between points and therefore reflects their dynamic connectivity in a diffusion process. This might be a subtle point leading to the differences in outcomes between the two reduction types that find a similar number of dominant directions.

%%^^^^^^^^ CHECK 

%%% OK vvvvvvvvvvvvvvvvvvvvvvvvv

We need to point out that the aforementioned analysis primarily focuses on the first two dominant eigenvectors and may not fully represent the inherent structure of multidimensional projections. Nonetheless, it enables us to make two assumptions. First, the density of points in the projections may vary with temperature across all considered descriptors and reduction methods. Second, the observed disparities between diffusion coordinates and principal components in scales, shapes, and scattering of points could be responsible for the different performances of DM and PCA in describing the data structures and their temperature dependencies.

%%%% ^^^^^^^^^^^^^^^^^^^^^^^^^^^^^

\subsection{Pattern of the glass transition}

\begin{figure}[tbhp]%[H]%[tbhp]
\centering
\includegraphics[width=1.0\columnwidth]{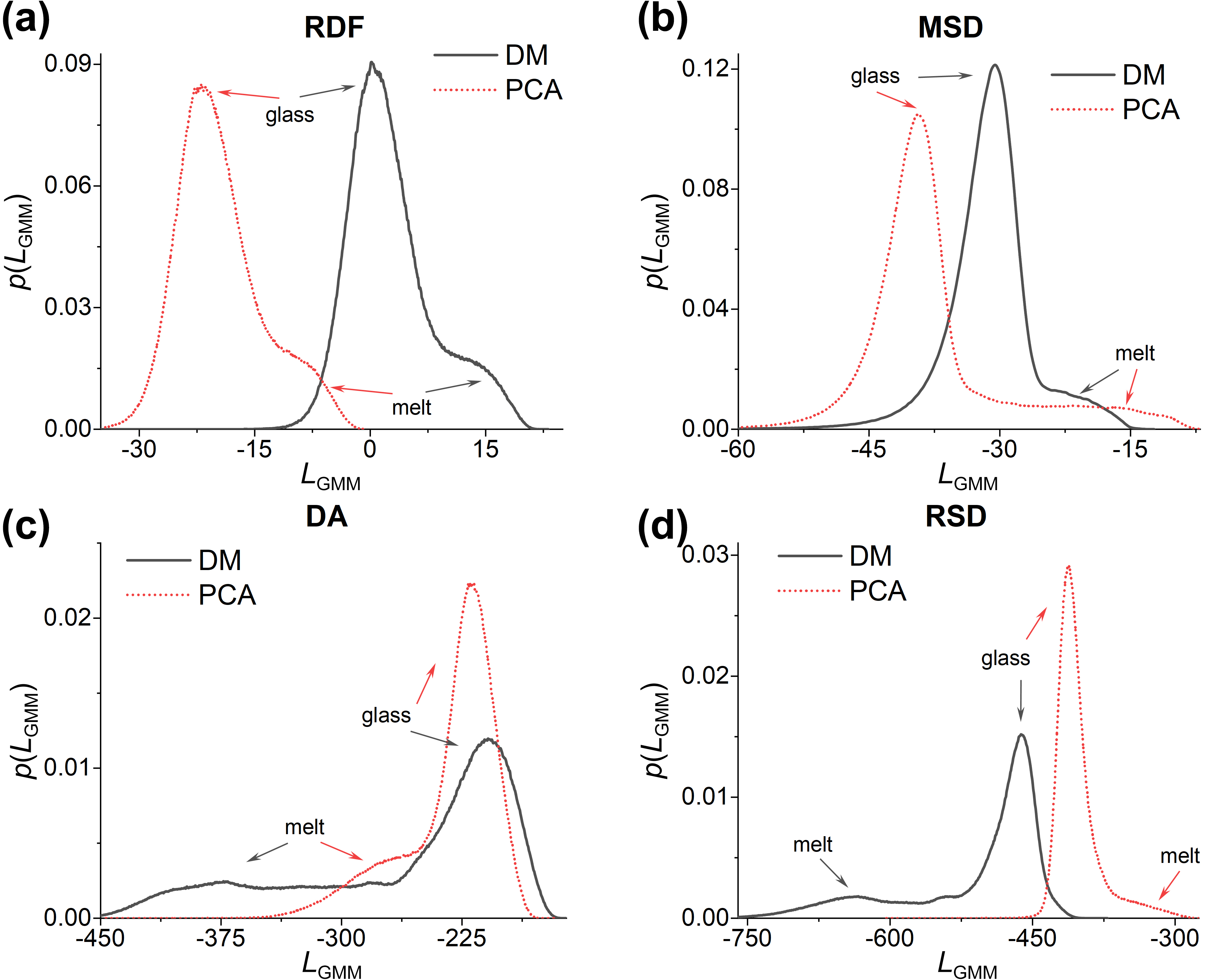}
\caption{Probability density distributions $p(L_\mathrm{GMM})$ of log-likelihoods $L_\mathrm{GMM}$ collected across the entire temperature scan and derived using GMMs of DM and PCA projections for the (a) RDF, (b) MSD, (c) DA, and (d) RSD data of PLA cooled after 3.5\,$\mu$s of simulations. Arrows indicate the melt and glass states identified depending on temperature, see 
Figure~S19. Results for PHB are shown in Figures~S20~and~S21.
}
\label{fig:p_LGMM}
\end{figure}

We now examine the temperature dependence of the densities of the DM and PCA projections. To address this point, we built a probabilistic representation of projections using a Gaussian Mixture Model. GMM models provide approximations of the probability density functions underlying the projections, which enable the use of log-likelihoods $L_\mathrm{GMM}$ as density-based metrics~\cite{McLachlan2019, Bishop_undated-lj}. A greater $L_\mathrm{GMM}$ value for a point indicates that it belongs to a neighborhood possessing a higher local density of projection. Having derived the $L_\mathrm{GMM}$ values (Section~\ref{sec:ml_methods}) for every projected point at each temperature, we evaluated how probable these metrics are and considered their distributions. In particular, we collected the log-likelihoods across the entire temperature scan and scrutinized their probability density distributions $p(L_\mathrm{GMM})$ for the different descriptors and reduction types in the case of PLA cooled after 3.5\,$\mu$s of simulations, Figure~\ref{fig:p_LGMM}. $p(L_\mathrm{GMM})$ at different $T$ are depicted separately in 
Figure~S19, 
serving as a basis to identify the high- and low-$T$ domains corresponding to the melt and glass states in Figure~\ref{fig:p_LGMM}, respectively. 
Results for PHB are presented in 
Figures~S20~and~S21.

Figure~\ref{fig:p_LGMM} shows that $p(L_\mathrm{GMM})$ curves exhibit complex shapes and for the most part share their common features.
%for different analyzed with DM and PCA. 
The most prominent feature is a pronounced peak, indicating the grouping of $L_\mathrm{GMM}$ from multiple $T$ into one population in the low-$T$ domain, 
Figure~S19. 
This suggests that the densities of the DM and PCA projections remain relatively unchanged upon cooling.
%of a glassy system. 
Furthermore, the peak is accompanied by a shoulder, or tail, of varying length for different descriptors of PLA and reduction methods, except for the RSD data transformed by DM. In the latter case, there is a less pronounced second peak at low $L_\mathrm{GMM}$ values, Figure~\ref{fig:p_LGMM}d. Interestingly, the high-$T$ peak is also observed for the DA data of PHB in the case of both DM and PCA due to a notable splitting of the two states, see 
Figure~S20c. 
These shoulders, tails, and peaks point to the presence of another distinctive population of $L_\mathrm{GMM}$ formed as a consequence of the significant influence of temperature on projection densities in the melt state 
(see Figures~S19~and~S21).

A closer examination of Figure~\ref{fig:p_LGMM}a reveals that the 
%$p(L_\mathrm{GMM})$ 
shapes of the distributions 
for
%in the case of 
the RDF data appear very similar for both DM and PCA.
The same extent of separation possibly suggests that the two reductions may capture the transition of $L_\mathrm{GMM}$ values from the melt-state population to the glass-state population at a similar temperature (Figures~S19a,b). 
The $p(L_\mathrm{GMM})$ curve from PCA is shifted toward lower $L_\mathrm{GMM}$ values corresponding to lower projection densities, in line with observations in Figure~\ref{fig:dm_and_pca_projections_of_rdf}a. In the case of MSD, Figure~\ref{fig:p_LGMM}b shows that the high-$T$ tail in $p(L_\mathrm{GMM})$ from DM is markedly shorter due to a weaker separation of the two $L_\mathrm{GMM}$ populations compared to PCA. This discrepancy may lead to differences in transition temperatures obtained using DM and PCA, with PCA potentially yielding a lower transition temperature as a greater number of temperatures display the $L_\mathrm{GMM}$ values inherent to the melt state 
(Figure~S19c,d). 
In addition, the $L_\mathrm{GMM}$ values at high $T$ are greater for PCA than for DM and vice versa at low $T$ (Figures~S19c,d), which contradicts the picture from 
Figure~S8 
and points to limitations of the corresponding analyses of the first two dominant directions. 

As far as the DA descriptor is concerned, Figure~\ref{fig:p_LGMM}c shows that a larger fraction of the $L_\mathrm{GMM}$ values forms a population corresponding to a melt state for DM compared to PCA. This indicates that, in terms of projection densities, the separation between the glass and melt states is stronger, and the transition temperature is probably lower for DM than for PCA (Figures~S19e,f). 

The RSD data transformed with DM clearly separate into two populations, see high-$T$ and low-$T$ peaks at low and high $L_\mathrm{GMM}$ values in Figure~\ref{fig:p_LGMM}d, respectively. In contrast, PCA leads to a weak separation and probably a high transition temperature (Figures~S19g,h). Moreover, the variation of $L_\mathrm{GMM}$ values with $T$ for PCA qualitatively differs from that for DM: a high-$T$ tail is located in a domain of high $L_\mathrm{GMM}$ values, see Figure~\ref{fig:p_LGMM}d. This means that cooling generally results in an increase of projection sizes, although the first two PCs show the opposite trend in 
Figures~S10c,d. 
Notably, the analysis of $p(L_\mathrm{GMM})$ at different $T$ 
(Figure~S19h) 
shows that $L_\mathrm{GMM}$ values decrease as the temperature decreases until about $T$ = 350\,K, after which they begin to slowly increase upon further cooling. Such a temperature behavior turns out to be specific only for the RSD descriptor reduced by PCA and is also observed for PHB, see 
Figures~S21. 
This underscores not only quantitative, but also qualitative differences in the performances of DM and PCA in the case of RSD.

Thus, the densities of DM and PCA projections
%for the considered MD data 
do depend on temperature. When collected over a cooling scan, they allow establishing a pattern of the glass transition: the separation of projection densities into two populations corresponding to the glass and melt states. These populations manifest themselves as high peaks in $p(L_\mathrm{GMM})$ for the glass state, together with tails, shoulders, or small peaks for the melt state. Visual analysis of $p(L_\mathrm{GMM})$ shapes indicates that the fraction of $L_\mathrm{GMM}$ values in each state depends on both the descriptor and transformation types. Therefore, various descriptors and reduction methods have different performance and thereby may lead to different $T_\mathrm{g}$ estimates.

\subsection{$T_\mathrm{g}$ determination}
\label{sec:determination}

According to the pattern derived in the previous section, and similarly to the dilatometric approach, the glass transition can be viewed as a continuous process of changing the projection densities under the influence of temperature. Since the extent of separation of the glass and melt states in terms of projection densities varies depending on the glass former, as well as the descriptor and reduction types, it is unclear what $L_\mathrm{GMM}$ value can be used to unambiguously differentiate the states in $p(L_\mathrm{GMM})$ (Figure~\ref{fig:p_LGMM}). Due to a weakly defined transition point, clustering of projections depending on a certain value of their density-based metric in the spirit of recent studies\cite{Banerjee2023-wh,Banerjee2023-xx} may lead to biased results. To avoid introducing any bias 
%transition point 
into our framework,
%by any means, 
we propose the following approach: the two-population pattern in $p(L_\mathrm{GMM})$ is formed because changes in projection densities weaken upon cooling, see 
Figure~S19. 
Therefore, one can expect variation in the overlap between $p(L_\mathrm{GMM})$ distributions at different temperatures. To quantify this overlap, we define an overlap parameter $O_\mathrm{p}$ at every temperature over the cooling scan as
\begin{equation}
O_p(T_i) = \sum_{\substack{j=1 \\j \ne i}}^{550} \int p(L_\mathrm{GMM}(T_i)) p(L_\mathrm{GMM}(T_j)) dL_\mathrm{GMM},    
\label{eq:O_p}
\end{equation}
%(1)
%
where $p(L_\mathrm{GMM}(T_{k}))$ denotes the probability density distribution of log-likelihoods at a temperature $T_{k} \,\,(k \in [1, 550])$.

\begin{figure}[tbhp]%[H]%[tbhp]
\centering
\includegraphics[width=1.0\columnwidth]{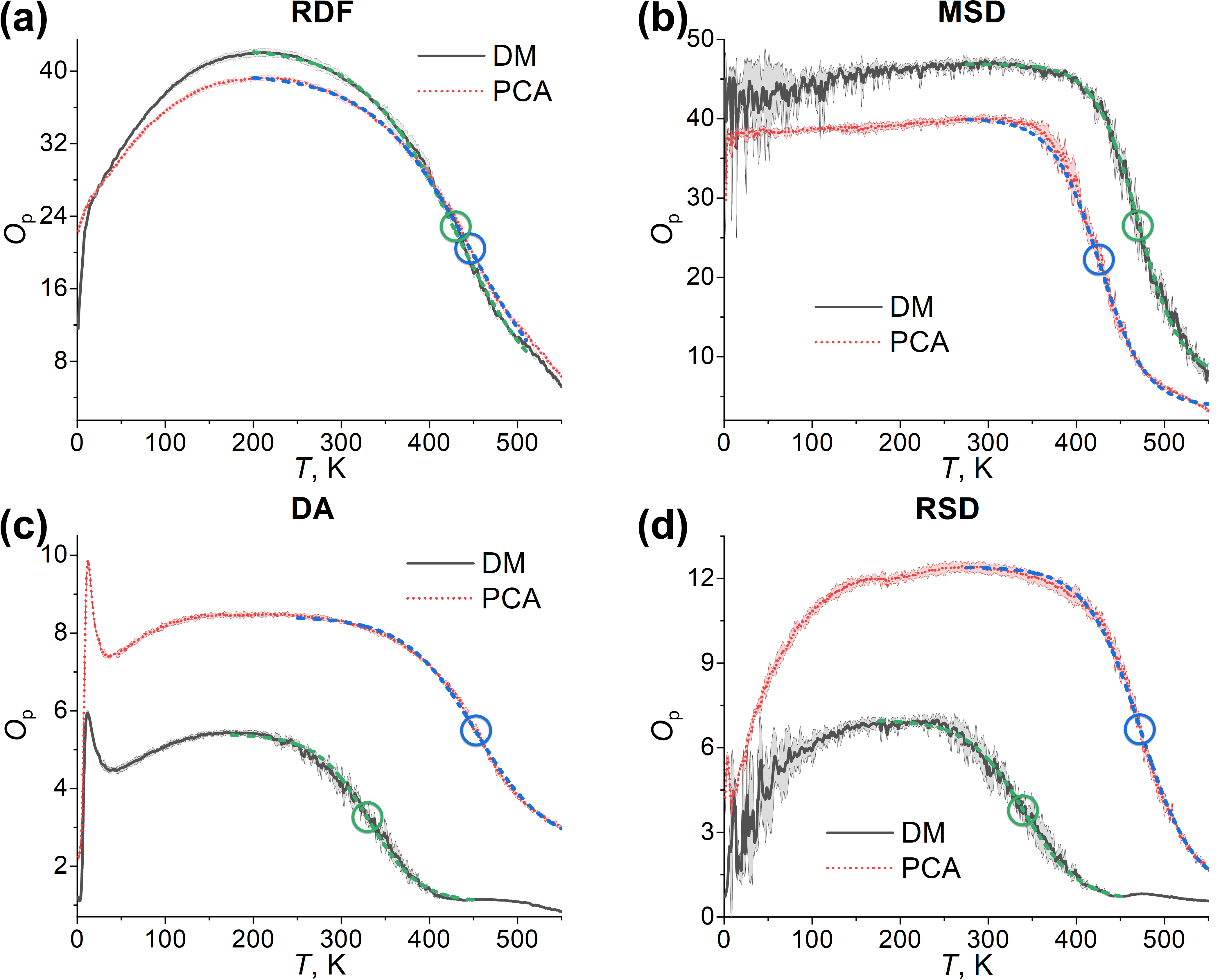}
\caption{Temperature dependence of the overlap parameter $O_\mathrm{p}$ (Equation~\ref{eq:O_p}) calculated based on $p(L_\mathrm{GMM})$ distributions at varied $T$ for DM and PCA projections of the (a) RDF, (b) MSD, (c) DA, and (d) RSD data of PLA. The dashed curves represent fits by using
Equation~\ref{eq:O_pT}, while circles point to $T_\mathrm{g}$s determined from the fits. 
Shaded regions indicate the standard deviation across three cooling scans. Results for PHB are shown in Figure~S22.
}
\label{fig:temp_dep_of_Op}
\end{figure}

The results obtained for the different descriptors from DM and PCA are presented in Figure~\ref{fig:temp_dep_of_Op} for PLA, and in 
Figure~S22 
for PHB. The figures show that at high $T$ the $O_\mathrm{p}$-values are lower relative to those at low $T$ of about 200\,K regardless of the descriptor and reduction type. This can be explained by the following facts: (i) $L_\mathrm{GMM}$ values in the melt state are notably influenced by temperature, and their distributions weakly overlap; (ii) they distinctly separate from those in the glass state, see 
Figure~S19. 
An abrupt increase in $O_\mathrm{p}$ becomes apparent at a certain $T$, reflecting the transition of the $L_\mathrm{GMM}$ values from the high-$T$ population to the low-$T$ population (Figure~\ref{fig:p_LGMM}). We thus propose that this increase serves as a hallmark of the glass transition. Then, the $O_\mathrm{p}$-values approach a plateau, the temperature range of which depends on the descriptor and reduction types, Figure~\ref{fig:temp_dep_of_Op}. The overlap is pronounced for the glass state, as the corresponding $p(L_\mathrm{GMM})$ curves tightly group with each other, see 
Figure~S19. 

However, variations in the values of $O_\mathrm{p}$ can occur upon further cooling, as seen for the RDF, DA, and RSD data in Figures~\ref{fig:temp_dep_of_Op}a,c~and~d at $T< 200$\,K, respectively. The shape of the $O_\mathrm{p}(T)$ 
curve
%dependencies 
in these cases turned out to be slightly different for PHB, see 
Figure~S22.

The study of 
%the domain of 
very low temperatures probably requires a dedicated analysis of much longer simulations, which is beyond the scope of this paper. Nevertheless, a distinct
%sharp 
change in $O_\mathrm{p}(T)$ is observable at $T$ above 200\,K in all cases considered. This facilitates the measurement of the 
$T_\mathrm{g}$ value using a hyperbolic tangent function\cite{Baglay2015}:
\begin{equation}
O_\mathrm{p}(T) = O_\mathrm{p}^\mathrm{av} + \frac{\Delta O_\mathrm{p}}{2} \tanh{\left(2 \frac{T_\mathrm{g}-T}{\omega} \right)} ,
\label{eq:O_pT}
\end{equation}
where $O_\mathrm{p}^\mathrm{av} = 0.5(O_\mathrm{p}^\mathrm{glass} + O_\mathrm{p}^\mathrm{melt})$ is the average $O_\mathrm{p}$-value at the beginning ($O_\mathrm{p}^\mathrm{melt}$) and end ($O_\mathrm{p}^\mathrm{glass}$) of the transition, $\Delta O_\mathrm{p}= O_\mathrm{p}^\mathrm{glass} - O_\mathrm{p}^\mathrm{melt}$ denotes the gap in $O_\mathrm{p}$-values for the two states, and $\omega$ describes the transition width.

Table~\ref{tab:DM_PCA} lists the resulting 
%ML 
$T_\mathrm{g}$ values for different descriptors and reduction types, alongside the dilatometric and dynamic $T_\mathrm{g}$ values 
(Figure~S2), 
as well as typical experimental data collected in Ref.~\cite{Van_de_Velde2002-vp} 
in the case of PLA and PHB. 
Concerning the RDF data of PLA, DM and PCA yield 
$T_\mathrm{g}$ values greater than the dilatometric $T_\mathrm{g}$ by about 9 and 25\,K, respectively. In the case of PHB and the same descriptor, DM outperforms PCA again and consistently gives a $T_\mathrm{g}$-value exceeding the dilatometric $T_\mathrm{g}$ by only about 13\,K, whereas PCA underestimates the transition point by about 25\,K. 
For the MSD data of PLA, DM gives the result perfectly coinciding with the dynamic $T_\mathrm{g}$, whereas PCA underestimates the transition temperature by about 43\,K. DM also outperforms PCA for the MSD descriptor of PHB: a dynamic $T_\mathrm{g}$ is underestimated by about 9 and 52\,K for DM and PCA, respectively. It is worth noting that the RDF- and MSD-based $T_\mathrm{g}$ values, along with their dilatometric and dynamic counterparts are
considerably higher than experimental $T_\mathrm{g}$s for the polymers under study. This outcome is typical and stems from the higher cooling rates in simulations compared to experimental ones\cite{Soldera2006-cd}.

\begin{table}
    \centering
    \begin{tabular}{|c|c|c|c|c|c|c|} \hline
     \multicolumn{1}{|c|}
     {\multirow{2}{*} 
     {\,} } 
     &  \multicolumn{2}{|c|} {$T_\mathrm{g}$PLA (K)}  
     & \multicolumn{2}{|c|} {$T_\mathrm{g}$PHB (K)} 
     &  \multicolumn{2}{|c|} {$T_\mathrm{g}$PLA-$T_\mathrm{g}$PHB (K)} 
     \\ \cline{2-7} %\hline
         & \textbf{DM} & \textbf{PCA} & \textbf{DM} & \textbf{PCA} & \textbf{DM}& \textbf{PCA}\\         \hline
      RDF  & $430 \pm 1$  & $446 \pm 2$ & $396 \pm 2$ & $358 \pm 4$ &  $34 \pm 2$ & $88 \pm 5$ \\ \hline 
      MSD  & $470 \pm 1$ &  $426 \pm 2$ & $437 \pm 1$ & $394 \pm 2$ & $33 \pm 1$ & $32 \pm 3$ \\ \hline
      DA  & $330 \pm 3$ & $453 \pm 2$ & $263 \pm 3$ & $356 \pm 1$ & $67 \pm 4$ & $97 \pm 2$\\ \hline
      RSD   &$340 \pm 6$  &$473 \pm 3$  & $301 \pm 4$ & $439 \pm 1$ & $39 \pm 7$ & $34 \pm 3$ \\ \hline
      $V(T)$   &  \multicolumn{2}{|c|}{421 ± 2}  &  \multicolumn{2}{|c|}{383 ± 1}  &  \multicolumn{2}{|c|}{38 ± 2} \\ \hline
      $\langle r^2(\Delta t^*) \rangle (T) $   &  \multicolumn{2}{|c|}{469 ± 2}  &  \multicolumn{2}{|c|}{446 ± 1}  &  \multicolumn{2}{|c|}{23 ± 2} \\ \hline
       experiment\cite{Van_de_Velde2002-vp}  &  \multicolumn{2}{|c|}{328–338} &  \multicolumn{2}{|c|}{278–288} &   \multicolumn{2}{|c|}{40–60} \\ \hline
    \end{tabular}
    \caption{The DM and PCA outcomes of $T_\mathrm{g}$ analysis, based on fitting the $O_\mathrm{p}$ curves by using Equation~\ref{eq:O_pT} for different descriptors of PLA and PHB, together with the dilatometric and dynamic $T_\mathrm{g}$ values 
    (Figure~S2) 
    and typical experimental data as reported by van de Velde \textit{et al.}\cite{Van_de_Velde2002-vp}.
The errors are the standard deviation across three cooling scans.}
    \label{tab:DM_PCA}
\end{table}

%When it comes to 
For the DA and RSD descriptors, $T_\mathrm{g}$ values obtained by DM of DA and RSD, surprisingly, are very close to experimental references for both polymers. In the case of DA, the simulation results match the experimental range perfectly for PLA, albeit underestimating the lower experimental values by only about 15\,K for PHB. The RSD-based $T_\mathrm{g}$ value perfectly agrees for PLA, while it is only about 13\,K higher for PHB compared to the corresponding upper experimental value. In contrast, PCA yields $T_\mathrm{g}$ values that significantly exceed the experimental values by about 115\,K and 135\,K in the case of DA and RSD for PLA, respectively. At the same time, PCA overestimates the PHB reference values by about 68\,K and 151\,K for DA and RSD, respectively. Thus, these two descriptors analyzed with DM give intriguing outcomes. The unexpected findings may be related to a weak dependence of these $T_\mathrm{g}$ values on the cooling rate. However, the verification of this hypothesis is a very resource-intensive task: an increase of the cooling rate by one order of magnitude, up to 1\,K/1\,ns, would require the production and analysis of about 3.5\,TB of simulations per scan (not accounting for the space required to store the descriptor data). Furthermore, obtaining various slow cooling rates for multiple configurations adds to the computational demand. Therefore, the present paper does not extend to this important yet challenging task.

In turn, the difference between the $T_\mathrm{g}$s of simulated PLA and PHB can be directly compared with experimental data\cite{Soldera2006-cd}. It is evident that the standard methods and both reduction techniques 
%accurately reproduce 
predict
the PLA transition to a glassy state at a temperature higher than that of PHB. However, the dynamic $T_\mathrm{g}$ underestimates the difference by about 17\,K compared to the lower reference value. 
%As for the $T_\mathrm{g}$s, 
The RDF and MSD data analyzed with DM lead to a slight underestimation by about 6 and 7\,K. This is also the case for the MSD and RSD data analyzed with PCA, where the difference in the $T_\mathrm{g}$s is about 8 and 6\,K lower, respectively. Moreover, PCA of the RDF and DA data yields notably overestimated results. Except these two cases, the quantitative differences in $T_\mathrm{g}$ values between the polymers from all the considered $T_\mathrm{g}$ estimates are in good agreement with experimental data.

Overall, DM and PCA applied to the RDF and MSD data tend to reproduce the results of dilatometric and dynamic methods, respectively. However, DM clearly outperforms PCA for these descriptors and gives consistent results for both PLA and PHB. As far as the DA and RSD data are concerned, the $T_\mathrm{g}$ values derived from DM are 
%unexpectedly close to 
in agreement with the
experimental references for the two polymers, whereas PCA tends to produce overestimated outcomes.

\subsection{Aspects of framework}

To better guide possible applications of dimensionality reduction to MD simulations, we will conclude with a brief discussion of several aspects related to the input MD data and ML protocol that may influence the $T_\mathrm{g}$ results. Among them are the selection of the writing step for MD trajectories, and the choice of observables considered. Moreover, we address the setting of the dimensionality of the projected space and test two alternative metrics for projection density.

\subsubsection{Writing step} 
\label{sec:writing_step}

%%% See the comment vvvvvvvvvvvvvvv

In this study, writing step of 25\,fs was selected based on the autocorrelation function for dihedral angles as outlined in Section~\ref{sec:methods}. Here, we additionally repeat our analysis for writing steps of 10\,fs, 100\,fs, and 1\,000\,fs for PLA. Note that all the data analyzed for the four different WS are obtained from the exact same trajectory file. The projection dimensionalities set at each WS are listed in 
Table~S2, 
while $T_\mathrm{g}$ results are presented in 
Table~S3. 

%%%^^^^^^^^^^^^^^^^^^^

Decreasing the WS from 25\,fs to 10\,fs did not change the outcomes. In turn, increasing the WS to 100\,fs affected the DA-based estimate from DM, leading to a higher $T_\mathrm{g}$ compared to WS\,=\,25\,fs by about 20\,K. For PCA, selecting WS\,=\,100\,fs resulted in a minor increase of RDF-based $T_\mathrm{g}$ and a slight decrease of the DA-based $T_\mathrm{g}$ compared to those at WS\,=\,25\,fs. A further increase of WS to 1\,000\,fs significantly impacted all the outcomes across the descriptors and reduction methods, except for the RSD data analyzed with PCA. Additionally, no agreement between the $T_\mathrm{g}$ data, and the dilatometric and dynamic results was found.

Thus, this additional analysis validates the results obtained at WS\,=\,25\,fs. It also highlights that care must be taken when selecting the writing step in MD simulations of glass formers in order to obtain robust outcomes and enable comparisons between different studies.

\subsubsection{Number of observables}
\label{sec:no_observables}

%%% vvvvvvvvvvvvvvv see the comments

Above, the main focus was on the analysis of RDF, MSD, and RSD data for chiral atoms, as well as on DAs for bonds connecting the chiral and the oxygen atoms of the ester groups. While this approach simplified computations, it did not encompass data from the other atoms and dihedral angles. We now investigate whether these findings change when the range of atom and dihedral angle types is expanded, see 
Figure~S4 
for the definitions of the additional observables included in this analysis. The dimensionalities of the projections
%estimated for 
from the extended datasets are presented in
Table~S4. 
The resulting temperature dependencies of $O_\mathrm{p}$ and $T_\mathrm{g}$ values for the PLA and PHB configurations cooled after the production runs are shown in 
Figures~S23 and Figures~S24, 
respectively. 

Figures~S23a and S24a 
show that considering most heavy atoms within the polymer chains slightly decreased the overestimated RDF-based $T_\mathrm{g}$ values from DM, aligning them closely with the dilatometric counterparts. For PCA of the RDF data, the extended datasets resulted in a more pronounced overestimation of $T_\mathrm{g}$s. In turn, additional atoms had practically no effect on the outcomes from the MSD and RSD data for both DM and PCA 
(Figures~S23b,f,d,h~and~S24b,f,d,h), 
except for the RSD data analyzed with PCA, where a difference of about 20\,K was seen (Figure~S24h). 
Interestingly, the extra DAs of PLA led to a 40\,K increase in the $T_\mathrm{g}$ value for DM, whereas about 73\,K decrease was observed for PCA 
(Figures~S24c,g). 
The increase can be explained by the addition of more restricted DAs, especially those associated with the plane ester group in the extended input dataset. Their transitions may require greater activation energies and subsequently the DAs freeze at higher temperatures. In this regard, DM provides outcomes that are more physical than the ones from PCA, which points to a better performance of DM. For the extra DAs in the case of PHB, the increase observed with DM was comparable to that for PLA, while PCA was less consistent and resulted in a significant overestimation of the $T_\mathrm{g}$ value, see 
Figures~S24c,g. 

Thus, the analysis of additional data for the RDF, MSD, and RSD descriptors shows that the number of observables can be reduced without significantly affecting the estimation of $T_\mathrm{g}$. However, the framework appears to be sensitive to the choice of the DA type. We believe that this feature offers a significant advantage for gaining better insights into the glass transition and for further developments of force fields. Aspects and the importance of dihedral parameterization have been discussed, for example, by Tolmachev \textit{et al.}~\cite{Tolmachev2020-rw, Lukasheva2022-kj} and Rusu \textit{et al.}~\cite{Rusu2020-sq}. Van der Wel~\cite{Van_der_Wel2021-pe} provides a discussion of some experimental aspects.

%%^^^^^^^^^^^^^^^

%3.5.3. 
\subsubsection{Projection dimensionality} 
\label{sec:proj_dimensionality}

In order to evaluate whether the choice of projection dimensionality impacts the $T_\mathrm{g}$ estimate, we additionally consider the “elbow” method for this choice:
%. In particular, 
one can set the dimensionality by identifying an “elbow” in the eigenvalue spectrum. Despite its apparent simplicity, this method can suffer from ambiguity due to the challenge of precisely defining the endpoint of the “elbow” and determining which part of the “elbow” should be used. This is demonstrated in 
Figure~S25, 
where several options are shown for 
selecting the number of dimensions.

For all descriptors, $T_\mathrm{g}$ was found to be sensitive to the projection dimensionality. Namely, RDF-based $T_\mathrm{g}$ values from DM first weakly decreased with an increasing number of dimensions and then slightly increased for high-dimensional projections, see 
Table~S5. 
This was also the case for PCA, although changes in $T_\mathrm{g}$ values were much more pronounced. Concerning MSD-based $T_\mathrm{g}$ values presented in
Table~S6, 
a weak decrease with increasing projection dimensionalities was found for DM, while PCA showed a significant decrease even when the number of dimensions is distant from the “elbow”. In the case of DA transformed with DM, the $T_\mathrm{g}$ values first slightly increased and then decreased as projection dimensionality increased, whereas $T_\mathrm{g}$ values from PCA markedly increased depending on projection dimensionality 
(Table~S7). 
RSD-based $T_\mathrm{g}$ values from DM behaved similar to those from the RDF data 
(Table~S8). 
However, there was no clear dependence observed for PCA: $T_\mathrm{g}$ values might both significantly decrease and increase with increasing projection dimensionality 
(Table~S8). 

Overall, this analysis shows that an appropriate choice of projection dimensionality is important for $T_\mathrm{g}$ evaluation. While a relatively weak dependence of outcomes is observed for the DM method, PCA proves to be highly sensitive to this choice. The “elbow” method lacks applicability for setting the number of dimensions since the resulting $T_\mathrm{g}$ values have a complex dependence on projection dimensionality and do not seem to correlate with the “elbow” position. This advocates for the 
%conventional 
definition based on spectral gaps used for this study (Section~\ref{sec:dimensionality_projections}).

\subsubsection{Density-based metrics} 
\label{sec:density_based_metrics}

The results in Section~\ref{sec:determination} showed that the description of MD data is influenced by the choice of the dimensionality reduction technique leading to variations in the $T_\mathrm{g}$ estimates. Even when a certain reduction yields a satisfactory projection, it is essential to assess if the choice of the density metric applied influences the results.

To assess the above, we computed core distances, see Section~\ref{sec:methods}. These distances are closely related to the hyperparameters of the popular DBSCAN algorithm\cite{Schubert2017-wt}, where one can tune the target number density of points by specifying the maximum distance for neighbor search and the number of points within the neighborhood of a point. We also calculated the mean square distances as an alternative metric of how tightly packed the points are around a given point, Section~\ref{sec:methods}. The use of core and mean square distances instead of log-likelihoods lead to the results presented in 
Tables~S9~and~S10, 
respectively. 

In the case of the core distances, RDF-based DMs slightly underestimated $T_\mathrm{g}$ for PLA, and reproduced it for PHB compared to the dilatometric references (Table~S9). 
At the same time, PCA slightly overestimated the $T_\mathrm{g}$ values for both polymers but reproduced the difference in their $T_\mathrm{g}$s. In the case of the MSD data, using core distances did not lead to a notable increase in $O_\mathrm{p}(T)$ due to a weak separation of $L_\mathrm{GMM}$ values at different $T$, thereby prohibiting $T_\mathrm{g}$ estimation for the considered polymers and reduction types. For the DA and RSD descriptors, the results from DM were underestimated compared to experimental data for both PLA and PHB. PCA led to overestimated $T_\mathrm{g}$ values, except for the DA data of PHB, where the outcome was close to the reference. Both DM and PCA resulted in an overestimated difference in $T_\mathrm{g}$s for the DA data of the two polymers, but reproduced their difference for the RSD data. 

Mean square distances 
(Table~S10) 
performed better than the core distances: DM and PCA reproduced the difference between the PLA and PHB $T_\mathrm{g}$s for the RDF data, although $T_\mathrm{g}$ values were lower than the dilatometric references. In addition, the mean square distances enabled the use of the MSD descriptor, leading to virtually no differences between the DM and PCA outcomes, but yielding results close to those from the log-likelihoods for DM. For the DA and RSD data transformed by DM, the mean square distances led to underestimated $T_\mathrm{g}$ values compared to the experimental ones. In the case of PCA, overestimation was observed for the RSD data. At the same time, the DA-based $T_\mathrm{g}$ matched the reference for PLA and underestimated it for PHB. The difference in $T_\mathrm{g}$s for the DA data of the two polymers was overestimated by PCA. 

Thus, the comparison of the outcomes from the different density-based metrics shows a significant dependence on the choice of the metric. Among the three density-based metrics analyzed, log-likelihoods lead to more consistent outcomes against references and therefore can be recommended for further use.

\section{Summary}

\begin{figure}[tbhp]%[H]%[tbhp]
\centering
\includegraphics[width=1.0\columnwidth]{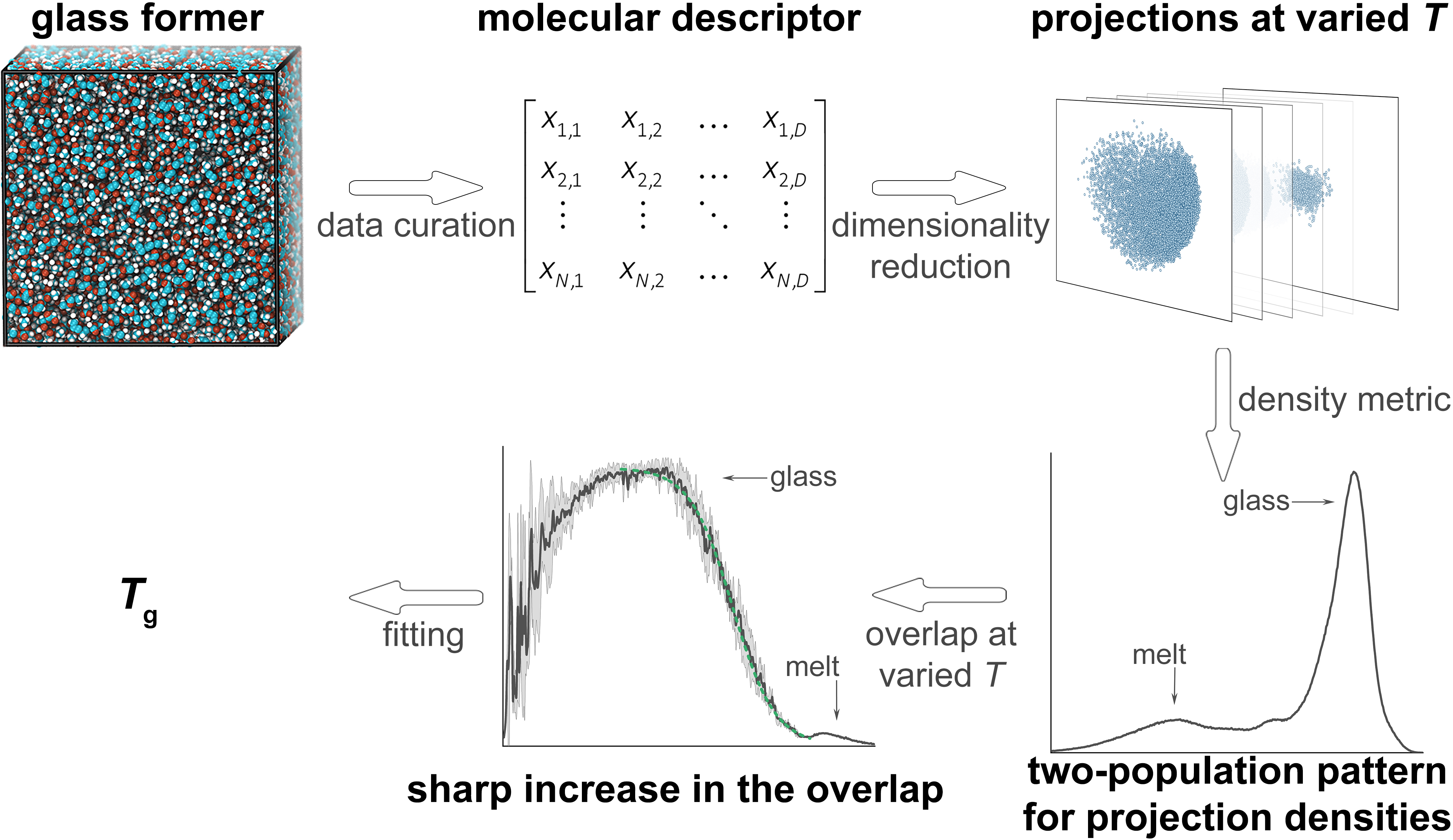}\\[10pt] 
\caption{Scheme of the joint MD and ML framework for evaluating $T_\mathrm{g}$ in six main steps: (1) MD simulations of a glass former; (2) collecting data for the molecular descriptors across the temperature scan; (3) dimensionality reduction to transform the descriptor data and obtain low-dimensional projections at varied $T$; (4) applying a density-based metric to the projections and calculating the distributions of the projection densities at each $T$ and across the entire scan; (5) if a two-population pattern is observed in the distribution based on all temperatures,
%of projection densities collected over all $T$, 
analyzing the overlap between the projection density distributions at varied $T$ can reveal the glass transition and facilitate estimating the $T_\mathrm{g}$ value by (6) fitting a sharp increase in the overlap.
}
\label{fig:scheme_of_MD_ML}
\end{figure}

We have shown that the densities of projections obtained through dimensionality reduction techniques correlate with the physical state of a system. Similar to the canonical determination of $T_\mathrm{g}$ based on volume or density analysis (the dilatometric approach), these projection densities are temperature-dependent and can group into two populations representing the melt and glass states. This observation enabled the development of a framework that combines computer simulations and machine learning methods to analyze various molecular descriptors and ultimately determine the corresponding transition temperatures in a unified manner, as schematically illustrated in Figure~\ref{fig:scheme_of_MD_ML}.

The framework was assessed based on cooling
%rate 
MD simulations of two polymers, PLA and PHB, for which the four molecular descriptors were analyzed: radial distribution functions, mean square displacements, dihedral angles, and relative square displacements. We considered principal component analysis and diffusion maps to obtain the projections, and analyzed their outcomes using three density-based metrics: log-likelihoods, core distances, and mean square distances. 

We put forward the conventional protocol for selecting the dimensionality of DM and PCA projections using spectral gaps and addressed the importance of this selection for the $T_\mathrm{g}$ analysis. When choosing log-likelihoods as the main metric, we found that both DM and PCA qualitatively reproduced the experimental difference in $T_\mathrm{g}$s for PLA and PHB, with PCA leading to a notable overestimation of the difference for the RDF and DA descriptors. Moreover, DM outperformed PCA and consistently produced the RDF- and MSD-based $T_\mathrm{g}$ values that align closely to their dilatometric and dynamic counterparts derived by the standard methods for both PLA and PHB. This affirms DM effectiveness and advocates for its continued utilization. 

Application of DM to the DA and RSD descriptors, the most pristine MD data, led to $T_\mathrm{g}$ values in agreement with experimental data, with deviations not exceeding 15\,K. For PCA, the discrepancy exceeded about 70\,K and more. We also showed that both DM and PCA outcomes are sensitive to the choice of the density-based metric: core distances and mean square distances turned out to be less robust compared to log-likelihoods. Furthermore, we proved that for a better predictive ability of the framework the writing step of simulation trajectories used to collect the descriptors data must be carefully chosen due to its influence on the resulting 
%ML 
$T_\mathrm{g}$ values. Taken together, we can recommend using the diffusion maps in conjunction with Gaussian Mixture Models to detect the glass transition and establish its temperature following the framework scheme in Figure~\ref{fig:scheme_of_MD_ML} for the considered molecular descriptors of various glass formers.

Our study unambiguously highlights the ability to observe the glass transition and determine the transition temperature based on mere microstructure data. We believe that the ability to resolve the physical states of systems with atomistic resolution using a density metric value for each projection point provides a valuable perspective to gain detailed insight into the glass transition within the bulk or near an interface. The framework presented herein is readily applicable to glass formers having a weakly defined transition range\cite{Lin2021-rm}. It can also be utilized not only for studying temperate transitions but also pressure transitions, when the dependence of glass former density on pressure is essentially non-linear across the scan\cite{Douglas2021-uv,Mukherjee2021-sa}. Finally, the framework is not limited to computer simulations as data sources, and it can be adapted for analyzing experimental data, such as radial distribution functions from scattering techniques\cite{Christiansen2020-de} and dihedral angles from nuclear magnetic resonance spectroscopy\cite{Van_der_Wel2021-pe}. This application is particularly intriguing because it may unlock new opportunities for the development of computational models directly based on experiments.

\section*{Conflicts of interest}
None to declare.

\section*{Supplementary Information}
%% We need to provide a description for each SI figure and table: Figure S1 shows X; Figure S2 shows Y; ... Table S1 lists XY...
Figure S1: time dependence of PLA and PHB chain dimensions; Figure S2: the system volume and mean square atomic displacements for PLA and PHB as functions of temperature; Figure S3: dihedral autocorrelation functions for PLA and PHB; Figure S4 defines atom and dihedral angle types for the extended datasets; Figure S5: radial distribution functions and mean square atomic displacements for PLA and PHB; Figure S6: dihedral angle distributions for PLA and PHB; Figure S7: double logarithmic plots of spectral gaps from PCA for various descriptors; Figure S8: the DM and PCA projections for the MSD data of PLA; Figure S9: the DM and PCA projections for the DA data of PLA; Figure S10: the DM and PCA projections for the RSD data of PLA; Figure S11: the DM and PCA projections for the RDF data of PHB; Figure S12: the DM and PCA projections for the MSD data of PHB; Figure S13: the DM and PCA projections for the DA data of PHB; Figure S14: the DM and PCA projections for the RSD data of PHB; Figure S15: the overlay of the DM and PCA projections for the RDF data of PLA; Figure S16: the overlay of the DM and PCA projections for the MSD data of PLA; Figure S17: the overlay of the DM and PCA projections for the DA data of PLA; Figure S18: the overlay of the DM and PCA projections for the RSD data of PLA; Figure S19: probability density distributions of log-likelihoods at different temperatures for various descriptors of PLA; Figure S20: probability density distributions of log-likelihoods based on all temperatures for various descriptors of PHB; Figure S21: probability density distributions of log-likelihoods at different temperatures for various descriptors of PHB; Figure S22: temperature dependence of the overlap parameter from DM and PCA for various descriptors of PHB; Figure S23: temperature dependence of the overlap parameter from DM and PCA for the extended datasets of various descriptors of PLA; Figure S24: temperature dependence of the overlap parameter from DM and PCA for the extended datasets of various descriptors of PHB; Figure S25 defines projection dimensionalities for various descriptors of PLA using the "elbow" method; Table S1 lists dimensionalities of the DM and PCA projections based on spectral gaps for various descriptors of PLA and PHB; Table S2 lists dimensionalities of the DM and PCA projections based on spectral gaps for various writing steps and descriptors of PLA; Table S3: the DM and PCA outcomes of $T_\mathrm{g}$ analysis depending on the writing step for various descriptors of PLA; Table S4 lists dimensionalities of the DM and PCA projections based on spectral gaps for the extended datasets of various descriptors of PLA; Table S5: the DM and PCA outcomes of $T_\mathrm{g}$ analysis depending on projection dimensionality for the RDF data of PLA; Table S6: the DM and PCA outcomes of $T_\mathrm{g}$ analysis depending on projection dimensionality for the MSD data of PLA; Table S7: the DM and PCA outcomes of $T_\mathrm{g}$ analysis depending on projection dimensionality for the DA data of PLA; Table S8: the DM and PCA outcomes of $T_\mathrm{g}$ analysis depending on projection dimensionality for the RSD data of PLA; Table S9: the DM and PCA outcomes of $T_\mathrm{g}$ analysis based on core distances for various descriptors of PLA and PHB; Table S10: the DM and PCA outcomes of $T_\mathrm{g}$ analysis based on mean square distances for various descriptors of PLA and PHB.

\section*{Data availability}

Data and parameters
are available at
\texttt{https://github.com/SoftSimu/glass\_transition}.

\section*{Acknowledgements}

MK thanks the Canada Research Chairs Program and the Natural Sciences and Engineering Research Council of Canada (NSERC) for financial support. The authors also thank the Digital Research Alliance of Canada for computational resources.

%\printbibliography

%\bibliography{bibliography}

\providecommand{\latin}[1]{#1}
\makeatletter
\providecommand{\doi}
  {\begingroup\let\do\@makeother\dospecials
  \catcode`\{=1 \catcode`\}=2 \doi@aux}
\providecommand{\doi@aux}[1]{\endgroup\texttt{#1}}
\makeatother
\providecommand*\mcitethebibliography{\thebibliography}
\csname @ifundefined\endcsname{endmcitethebibliography}  {\let\endmcitethebibliography\endthebibliography}{}

\end {document}